\documentclass[twocolumn,times]{aastex63}
\usepackage{hyperref}
\usepackage{physics}
\hypersetup{colorlinks, citecolor=blue, linkcolor=blue, urlcolor=blue}

\usepackage{comment}

\usepackage{CJK}

\received{December 11, 2020}
\submitjournal{AAS Journals}

\shorttitle{Boundary Layer Circumplanetary Accretion}
\shortauthors{Dong, Jiang, \& Armitage}
\graphicspath{{./}{figures/}}

\begin{document}
\begin{CJK*}{UTF8}{gbsn}

\title{Boundary Layer Circumplanetary Accretion: How Fast Could an Unmagnetized Planet Spin Up Through Its Disk?}

\correspondingauthor{Jiayin Dong}
\email{jdong@psu.edu}

\author[0000-0002-3610-6953]{Jiayin Dong (董 佳音)}
\altaffiliation{CCA Pre-Doctoral Fellow}
\affiliation{Department of Astronomy \& Astrophysics, The Pennsylvania State University, University Park, PA 16802, USA}
\affiliation{Center for Exoplanets \& Habitable Worlds, 525 Davey Laboratory, The Pennsylvania State University, University Park, PA 16802, USA}
\affiliation{Center for Computational Astrophysics, Flatiron Institute, 162 Fifth Avenue, New York, NY 10010, USA}

\author[0000-0002-2624-3399]{Yan-Fei Jiang (姜 燕飞)}
\affiliation{Center for Computational Astrophysics, Flatiron Institute, 162 Fifth Avenue, New York, NY 10010, USA}

\author[0000-0001-5032-1396]{Philip J. Armitage}
\affiliation{Center for Computational Astrophysics, Flatiron Institute, 162 Fifth Avenue, New York, NY 10010, USA}
\affiliation{Department of Physics and Astronomy, Stony Brook University, Stony Brook, NY 11794, USA}

\begin{abstract}
Gas giant planets are expected to accrete most of their mass via a circumplanetary disk. If the planet is unmagnetized and initially slowly rotating, it will accrete gas via a radially narrow boundary layer and rapidly spin up. Radial broadening of the boundary layer as the planet spins up reduces the specific angular momentum of accreted gas, allowing the planet to find a terminal rotation rate short of the breakup rate. Here, we use axisymmetric viscous hydrodynamic simulations to quantify the terminal rotation rate of planets accreting from their circumplanetary disks. For an isothermal planet-disk system with a disk scale height $h/r =0.1$ near the planetary surface, spin up switches to spin down at between 70\% and 80\% of the planet's breakup angular velocity. In a qualitative difference from vertically-averaged models---where spin down can co-exist with mass accretion---we observe \emph{decretion} accompanying solutions where angular momentum is being lost. The critical spin rate depends upon the disk thickness near the planet. For an isothermal system with a disk scale height of $h/r = 0.15$ near the planet, the critical spin rate drops to between 60\% and 70\% of the planet's breakup angular velocity. In the disk outside the boundary layer, we identify meridional circulation flows, which are unsteady and instantaneously asymmetric across the mid-plane. The simulated flows are strong enough to vertically redistribute solid material in early-stage satellite formation. We discuss how extrasolar planetary rotation measurements, when combined with spectroscopic and variability studies of protoplanets with circumplanetary disks, could determine the role of magnetic and non-magnetic processes in setting giant planet spins.
\end{abstract}

\keywords{hydrodynamical simulations (767)---planet formation (1241)---accretion (14)}

\section{Introduction} \label{sec:intro}
Planetary spin is one of the fingerprints of planet formation. In the absence of planetary winds or strong tidal effects, planetary angular momentum is conserved since the end of planet formation, providing direct evidence by which we can compare theoretical models of giant planet formation against observations. In the Solar System, Jupiter has a rotation period of $2\pi/\Omega_{\rm J}=9.9$~hours while that of Saturn (which is harder to measure) is estimated to be $2\pi/\Omega_{\rm S}=10.7$~hours. Compared to the nominal breakup angular velocity, which for a planet of mass $M_{\rm p}$ and radius $r_{\rm p}$ is $\Omega_{\rm p} = \sqrt{GM_{\rm p} / r_{\rm p}^3}$, $\Omega_{\rm J} / \Omega_{\rm b} = 0.30$ and $\Omega_{\rm S} / \Omega_{\rm b} = 0.39$. Extrasolar planetary spins measurements are in their infancy but can be inferred, in principle, from transit constraints on oblateness \citep{seager02,zhu14,biersteker17}, from combined photometric and spectroscopic data during transit \citep{akinsanmi20}, and from high resolution spectroscopy \citep{bryan18}. For five objects with masses at the upper end of the planetary mass range, \citet{bryan18} inferred rotation rates of 5-30\% of breakup, similar to those of a sample of brown dwarfs. 

Gas giant planets are predicted to accrete most of their mass via a circumplanetary disk \citep{miki82,dangelo03,papaloizou05,machida10,szulagyi14}. If the planet is unmagnetized and initially slowly rotating, it will accrete gas from its circumplanetary disk via a radially narrow boundary layer with specific angular momentum roughly equal to the Keplerian value at the planetary surface \citep{pringle77}. Rapid spin-up will ensue. Qualitatively different evolution occurs if the planet is sufficiently magnetized, with accretion occuring via a magnetosphere. Magnetic coupling between the planet and its circumplanetary disk can then regulate the spin to values $\Omega \ll \Omega_{\rm b}$. Disk braking has been studied in the context of Classical T~Tauri stars \citep{konigl91,edwards93,armitage96}, and would operate efficiently for Jupiter mass planets given an ordered surface field of $B \sim 500\,{\rm G}$ \citep{batygin18,ginzburg20}. The strength of the dipole component of Jupiter's current magnetic field is only about 4~G \citep{smith74}, though scaling arguments suggest that proto-Jupiters would generate much stronger fields \citep{christensen09}. By analogy with the stellar case, strong magnetic fields could be (relatively) directly detected through spectroscopic signatures of magnetospheric accretion \citep{edwards94}, or by observation of protoplanetary jets \citep{gressel13}.

\begin{figure}
    \centering
    \hspace*{0.1cm}
    \includegraphics[scale=1.15]{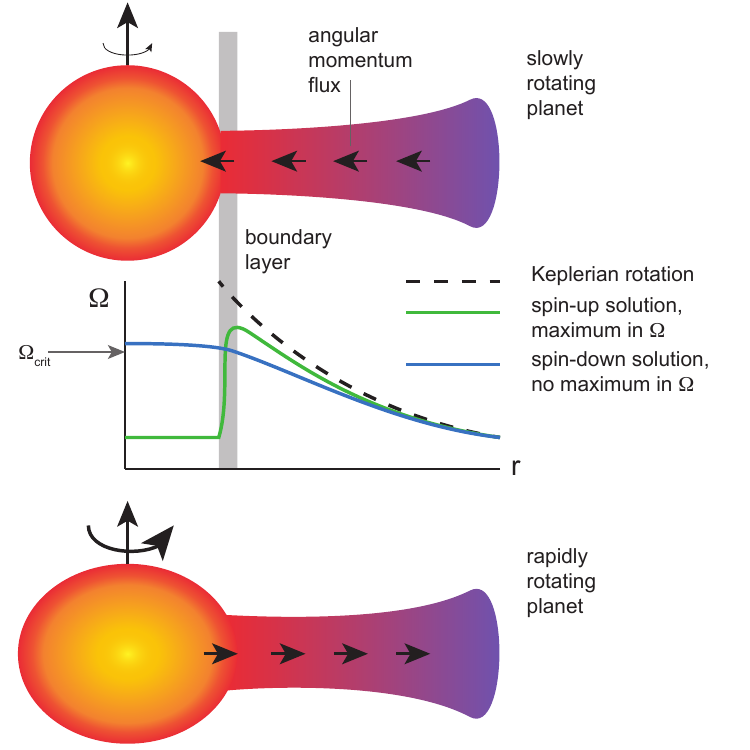}
    \caption{A schematic illustration of the interaction circumplanetary disk accretion occurring via a boundary later. For a slowly rotating planet (upper panel), the boundary layer is narrow, and the planet accretes gas whose specific angular momentum is approximately that of a disk orbiting at the planetary surface. For a rapidly rotating planet (lower panel), the boundary layer is broad, and there is no maximum in the angular velocity profile. Angular momentum transport, if modeled as a fluid viscosity, then works to transport angular momentum from the planet, through the boundary layer, and into the disk. Above some critical spin rate $\Omega_{\mathrm{crit}}$, the planet will spin down instead of spin up.}
    \label{fig:cartoon}
\end{figure}

Although a strong case can be made for magnetic regulation of planetary spin, the strength and ubiquity of protoplanetary magnetic fields remain uncertain. With that in mind, we focus on a simpler question: what is the terminal rotation rate of gas giant planets in the {\em unmagnetized} limit? One firm limit is the onset of bar instability, which occurs when the ratio of rotational kinetic energy to gravitational energy, $T / |W| \gtrsim 0.27$ \citep{durisen86}. A second limit arises from the hydrodynamics of boundary layer accretion. \citet{popham91}, using steady one-dimensional models of the disk and boundary layer, showed that the specific angular momentum of accreted gas equals that of the central object at a critical rotation speed that is {\em below} the breakup velocity. This behavior occurs because the boundary layer (defined as the region of the flow where $\Omega(r)$ is an increasing function of radius) becomes first broader with increasing spin, before ceasing to exist at the critical rotation speed. At and above the critical rotation speed, $\Omega(r)$ is a monotonically decreasing function, and angular momentum can be lost from the central object and transported through the disk to large distances. Figure~\ref{fig:cartoon} illustrates this transition.

In this paper, we calculate the terminal planetary spin rate for a model system in which an unmagnetized giant planet interacts viscously with a circumplanetary disk via a boundary layer. Following \citet{hertfelder17}, who simulated compact object boundary layers, we extend the work of \citet{popham91} in two ways. First, we simulate the planet-disk interaction in two-dimensions (assuming axisymmetry), rather than adopting a one-dimensional vertically averaged treatment. Axisymmetry allows for a better representation of the problem geometry, and captures meridional flows that can develop in viscous disks \citep{urpin84,philippov17}. Second, we evolve a time-dependent simulation toward a (quasi)-steady state, rather than solving directly for the time-independent hydrodynamic flow. This changes the mathematically character of the problem, allowing us in particular to explore the possibility that the solution changes from accretion to \emph{decretion} \citep{pringle91} as $\Omega_{\rm p}$ increases.

Multiple physical processes are important in boundary layers, and our model system, by construction, excludes some of them. In particular, the physical origin of angular momentum transport may differ between the boundary layer and the disk, because the magnetorotational instability \citep{balbus98} only operates where the angular velocity is a decreasing function of radius \citep{armitage02,steinacker02,pessah12,hertfelder15,philippov16,belyaev18}. We also ignore thermal effects. Around a slowly rotating central object, half of the total accretion energy is released in the boundary layer region, yielding a large luminosity that can modify the structure of the boundary layer and inner disk \citep{kley96}. Finally, our focus is on the consequences of boundary layer accretion for planetary spin evolution, and we do not study the impact of the accreted gas on the outer planetary envelope \citep{balsara09}.

The structure of the paper is as follows. In \S\ref{sec:physical} we describe how the simulation parameters, including the disk aspect ratio, envelope sound speed, and kinematic viscosity, map to the physical parameters of circumplanetary systems. \S\ref{sec:numerical} details the problem setup within the $\mathtt{Athena}$++ code \citep{stone20}. Our results are presented in \S\ref{sec:results}. We discuss implications of our findings, and avenues for further investigation, in \S\ref{sec:conclusion}.

\section{Physical Parameters} \label{sec:physical}
Our simulation parameters, including the disk aspect ratio, envelope sound speed, and kinematic viscosity, are motivated by physical conditions of the circumplanetary system. In this section, we justify the use of these parameters. 

The temperature of the inner circumplanetary disk and boundary layer are set by the mass accretion rate and the radial extent of the boundary layer. 
The mass accretion rate onto the planet through its boundary layer can be roughly estimated from the planet's mass and the circumplanetary disk lifetime. Assuming that circumplanetary disks have a comparable lifetime to protoplanetary disks \citep[i.e. a few Myr,][]{haisch01}, the mass accretion rate of the disk is $10^{-9}$--$10^{-8}M_\sun$\,yr$^{-1}$ for giant planets of a few Jupiter masses. Direct imaging of planets and their disks allows for limited observational constraints on the mass accretion rate. The H$\alpha$ emission from the forming planet PDS~70b, observed by the VLT/MUSE, constrains the planet's accretion rate to be of the order of $10^{-8}$--$10^{-7}M_\sun$\,yr$^{-1}$ \citep[e.g., ][]{wagner18}. Based on these considerations, we assume an accretion rate of $10^{-8}M_\sun$\,yr$^{-1}$ in our calculations. The protoplanetary radius can be estimated using theoretical models of early giant planet evolution. \citet{fortney11}, for example, find that Jupiter would have had a size of 1.6~R$_{\mathrm{Jup}}$ at an age of 1~Myr. With these values in hand, we can estimate the temperature of a steady-state circumplanetary disk $T_{\mathrm{disk}} (r)$ in the vicinity of the planet as \citep[e.g., ][]{armitage07},
\begin{equation}
    T_{\rm disk}^4 = \frac{3GM_{\mathrm{p}}\dot{M}_{\mathrm{BL}}}{8\pi \sigma r^3}\bigg(1-\sqrt{\frac{r_{\mathrm{p}}}{r}}\bigg),
\end{equation}
where $M_{\mathrm{p}}$ is the planet mass, $\dot{M}_{\mathrm{BL}}$ is the mass accretion rate onto the planet through the disk and boundary layer, $\sigma$ is the Stefan-Boltzmann constant, $r$ is the radial distance, and $r_{\mathrm{p}}$ is the planet size. For a one Jupiter-mass planet, with a radius of 1.6 Jupiter-radii, and an accretion rate of $10^{-8}M_\sun$\,yr$^{-1}$, we obtain a disk temperature at $r = 2 R_{\mathrm{Jup}}$ of $\sim$ 1600 K. The corresponding sound speed is, 
\begin{equation}
    c_s = \Big(\frac{kT_{\mathrm{disk}}}{\mu m_\mathrm{H}}\Big)^{1/2},
\end{equation}
where $\mu$ is the mean molecular weight. Assuming $\mu = 2.4$, ratio of the sound speed to the local Keplerian velocity is $c_s/v_{\mathrm{Kep}} \simeq 0.1$. For a circumplanetary disk in vertical hydrostatic equilibrium the disk aspect ratio is also $h/r = c_s/v_{\mathrm{Kep}} \simeq 0.1$.

For reasons of computational simplicity, we assume that the outer envelope of the planet, the boundary layer, and the circumplanetary disk are all isothermal. The structure of the boundary layer is expected to depend most strongly on the aspect ratio of the adjacent disk, so we set the sound speed so that $h/r$ would equal 0.1 if the disk extended to the planetary surface. The disk flares as $h/r \propto r^{1/2}$ at larger radii. The assumed sound speed implies a high but not grossly unreasonable temperature and pressure scale height in the outer envelope of the planet.

The physical origin of angular momentum transport in circumplanetary disks is yet not well understood. Several mechanisms are possible, including the magnetorotational instability \citep{balbus98} and spiral density waves \citep{zhu16}. We model angular momentum transport in the fluid system using a fixed kinematic viscosity $\nu$. The value of $\nu$ is set based on physical considerations \citep[primarily estimates derived from modeling of dwarf nova systems;][]{king07}, together with numerical constraints from the need to reach a time-independent accretion state throughout at least the inner disk. Setting $\nu = 10^{-3}$, which corresponds to a Shakura-Sunyaev $\alpha \sim 0.1$ in the inner disk, satisfies these requirements.

\section{Numerical Setup} \label{sec:numerical}
We perform 2-dimensional, axisymmetric viscous hydrodynamic simulations to study boundary layer circumplanetary accretion at different planetary spin rates using the $\mathtt{Athena}$++ \citep{stone20}  grid-based code. The equations of hydrodynamics solved in the code for our problem are,
\begin{gather}
    \pdv{\rho}{t} + \div{(\rho \vb*{v})} = 0,\\
    \pdv{(\rho \vb*{v})}{t} + \div{(\rho \vb*{v}\vb*{v} + P\,\vb{I} - \vb*{T})} = -\rho \grad{\phi_{\mathrm{p}}},
\end{gather}
where $\rho$ is the gas density, $\vb*{v}$ is the velocity vector, $P$ is the gas pressure written as $P = c_s^2 \rho$ for an isothermal equation of state, $\vb{I}$ is the identity tensor, $\vb*{T}$ is the viscous stress tensor, and $\phi_{\mathrm{p}}$ is a static gravitational potential from the planet. The mass of the planet exterior to the inner boundary is small, and so it suffices to treat the gravitational potential of the planet as a point mass at the origin. Since we restrict consideration to an isothermal and unmagnetized regime, the energy density and magnetic field equations are not evolved. 
We adopt a spherical-polar coordinate system ($r, \theta, \phi$) to achieve an axisymmetric setup in $\phi$-direction. The fiducial simulation domain spans $r \in [0.9, 10]$ and $\theta \in [0, \pi]$, with $N_r \times N_\theta = 2048 \times 2048$. We run one simulation in a larger domain ($r \in [0.9, 20]$), with identical spatial resolution, to test whether the flow features of interest are influenced by the outer boundary condition. In the $r$-direction, the planet is initialized at radii between 0.9 to 1.1, while the disk is initialized at radii between 1.1 to 10. We adopt a logarithmic grid setup with a ratio of 1.001176 (i.e., $r_{i+1} = 1.001176 r_{i}$). The setup allows us to resolve the planetary envelope and the boundary layer at high resolution (about 10 grid points per pressure scale height $h_{\mathrm{planet}}$ of the planet, where $h_{\mathrm{planet}} = c_s^2/g$ with an isothermal sound speed of 0.1), and resolve the outer disk region where the scale height is much larger ($h_{\mathrm{disk}}= c_s/\Omega$) at lower resolution to reduce computational cost. In the $\theta$-direction we use a uniform grid spacing. 
We use the $\mathtt{HLLE}$ Riemann solver with second order reconstruction for hydrodynamics.\footnote{We find the $\mathtt{Roe}$ Riemann solver leads to numerical divergences using our simulation setup. Compared to the $\mathtt{HLLE}$ Riemann solver, the $\mathtt{Roe}$ solver is more accurate but less diffusive and robust. The sharp density transition from the planetary envelope to the low density region near the pole  may cause the numerical issue with the $\mathtt{Roe}$ solver.} The number of ghost cells is set to be 4. Default $\mathtt{Athena}$++ configurations are used if not otherwise stated.

\begin{figure*}
    \centering
    \hspace*{0.3cm}
    \includegraphics[scale=1.02]{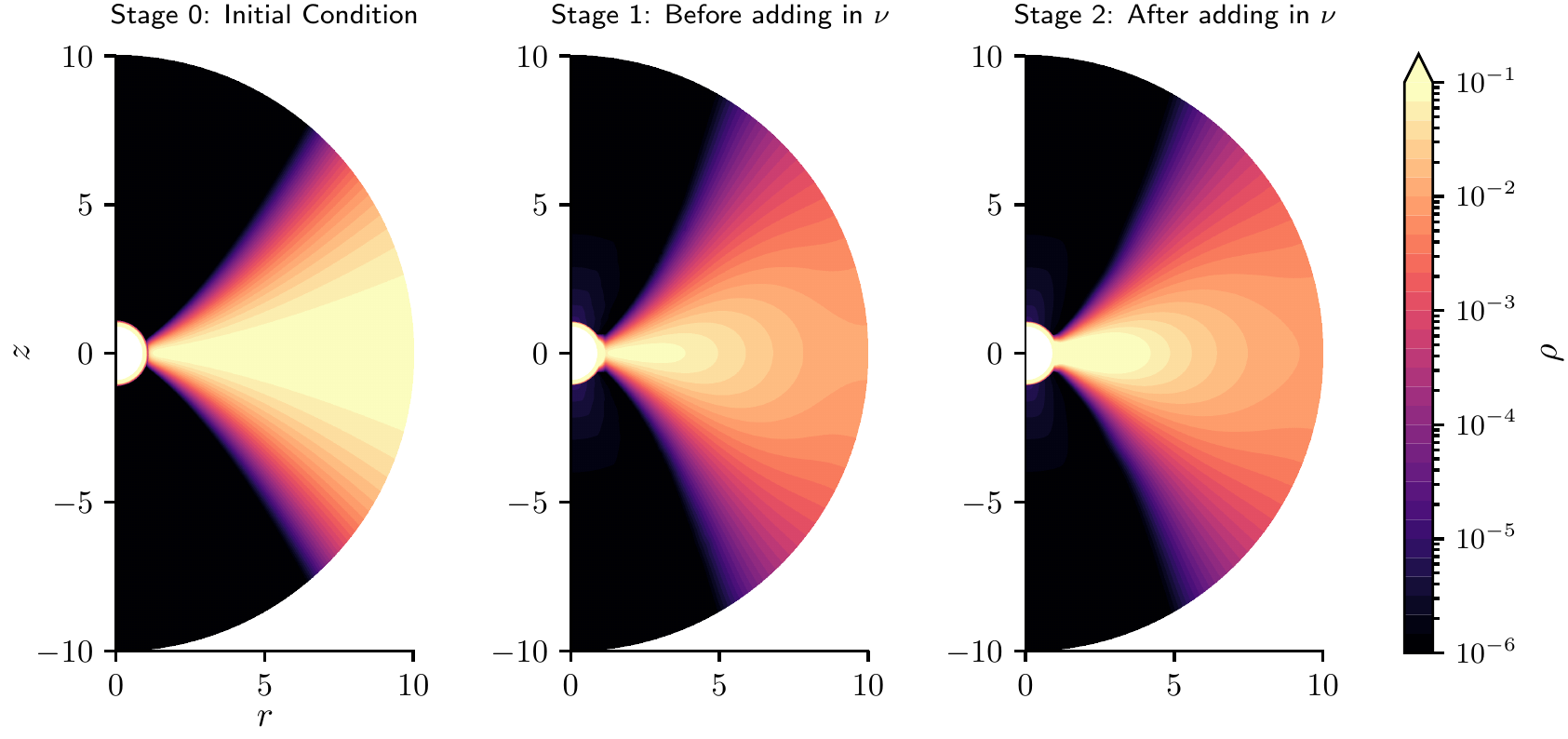}
    \caption{Illustration our two stages approach to the simulations. The left panel shows the density in the Stage 1 initial condition in which the disk is in vertical hydrostatic equilibrium, and there is zero planetary spin and zero viscosity. The middle panel shows the density after Stage 1 of the simulation with the central planet spun up to the desired angular velocity but the disk still having zero viscosity. The right panel shows the density profile after Stage 2 of the simulation with the viscosity $\nu$ added. The circumplanetary system has reached a quasi-steady state for the planet, boundary layer, and inner disk region.}
    \label{fig:density}
\end{figure*}

The planetary envelope is setup in a non-rotating hydrostatic equilibrium (satisfying $c_s^2\dv*{\rho}{r} = -\rho GM_{\mathrm{p}}/r^2$) with the following density profile as the initial condition,
\begin{equation}
    \label{eqn:hydro_planet}
    \rho(r) = \rho_0 \exp \left[ {\beta (r_{\rm p}/r-1)} \right],
\end{equation}
where $\beta = G M_{\mathrm{p}}/c_s^2 r_{\rm p}$ and $r \in [0.9, 1.1]$. At $r = r_{\rm p}$, $\rho = \rho_0$. With an isothermal equation of state, the planet does not have a sharply defined surface, but the boundary layer--planet system evolves to a steady-state in which it is reasonable to approximately identify $r_{\rm p}$ with the planetary ``surface". We set $G M_{\mathrm{p}}$, $r_{\rm p}$, and $\rho_0$ to unity. An isothermal sound speed $c_s = 0.1$ is used, as justified in \S\ref{sec:physical}. 

The circumplanetary disk is initialized in vertical hydrostatic equilibrium. The density profile of the disk in cylindrical coordinates with $R = r \sin{\theta}$ and $z = r \cos{\theta}$ is,
\begin{equation}
    \rho(R, z) = \Sigma_0 \exp \left[ {-z^2/2h_\mathrm{disk}^2} \right],
\end{equation}
where $\Sigma_0$ is the mid-plane density and $h_{\mathrm{disk}}$ is the vertical disk scale height. We set $\Sigma_0 = 0.1$ as a constant throughout the disk. The disk scale height, $h_{\mathrm{disk}}= c_s/\Omega = c_s r^{3/2}$, is an increasing function of the radial distance. We apply a density floor \emph{dfloor} $= 10^{-6}$ to avoid cells approaching zero density at large $z$. The disk initially rotates at the Keplerian angular velocity and we have $v_\phi = R^{-1/2}$.

As boundary conditions, we apply the polar-wedge boundary condition for the $\theta$ boundaries and  periodic boundary conditions for the $\phi$ boundaries. For the inner $r$ boundary, we extrapolate the hydrostatic equilibrium equation of the planet (Eq.~\ref{eqn:hydro_planet}) to the ghost cells and use a reflecting radial momentum boundary condition. For the outer $r$ boundary, we use an outflow boundary condition but set the radial velocity component of ghost cells to zero if it points inward to avoid infall.

Generating a numerically stable model of a rapidly rotating planet, as an effective initial condition for our simulations, requires some care. After experimentation, we have found it best to run the simulations in two stages, of which the first is driven spin-up of initially non-rotating planet models. Figure~\ref{fig:density} shows the scheme. In the left panel of Figure~\ref{fig:density}, we show the initial condition of the system with both the planet and the disk in hydrostatic equilibrium and the disk rotating at Keplerian angular velocity. The planet starts with zero angular velocity. During Stage 1 of our simulations, we gradually spin up the planet by adding a small amount of angular momentum to the planetary envelope at each timestep until it reaches the desired spin rate. The kinematic viscosity is set to zero at this stage. Since we are only interested in the interaction of the outer planetary envelope with the disk through the boundary layer, and spinning up the inner planetary envelope tends to numerically destabilize the inner radial boundary, we design an angular velocity profile that keeps the inner planetary envelope rotation at zero while the outer envelope rotates at the desired angular velocity. The angular velocity profile follows a logistic function, written as,
\begin{equation}
    \Omega(r, \theta) = \frac{\Omega_{\mathrm{p}} \sin{\theta}}{1+\exp\left[{-400(r-0.93)}\right]},
\end{equation}
where $\Omega_{\mathrm{p}}$ is the targeted angular velocity. We add an extra $\sin{\theta}$ term to the angular velocity profile to avoid planetary rotation near the $\theta$ boundaries that would otherwise lead to instability. Since the disk scale height is relatively small at the boundary layer ($h_{\mathrm{disk}} \sim 0.1$ at $r = 1$), the $\sin{\theta}$ term does not significantly reduce the planetary rotation at latitudes where the planet and the disk interact. As the planet gradually spins up, its density structure adjusts such that the pressure gradient and rotation jointly balance gravity. Therefore, for higher rotation rates, shallower pressure gradients are obtained. 
We add a small amount of angular momentum at each timestep such that over the time interval of 100-orbit at $r=1$, the angular velocity is increased by $\sim$ 0.1. We only modify the planetary angular velocity at $r \leq 0.96$ to make sure that we do not influence the physical behavior of the boundary layer. To avoid angular velocity overshooting, we also include a damping term which reduces the angular velocity by 0.0015 per timestep (i.e., corresponding to a damping timescale of 100-orbit at $r=1$) if it is above the designed profile. We run the first stage for 800~orbits at $r=1$ to make sure that the planetary envelope spins up to the expected angular velocity and the planet and circumplanetary system reach a steady state. Five sets of simulations are performed with five different planetary spin rates, $\Omega_0 \in [0., 0.3, 0.5, 0.7, 0.8]$. The snapshot of one of the circumplanetary systems after the initial 800-orbit integration is shown in the middle panel of Figure~\ref{fig:density}.

For Stage 2 of our simulations, we use the output profiles from Stage 1 as the initial conditions. We now add viscosity to the system at radii $r > 0.96$. We set the kinematic viscosity $\nu = \alpha c_s^2/ \Omega = 10^{-3}$ as a constant. This results in a radially dependent value of the effective $\alpha$ parameter \citep{shakura73}. Near the boundary layer, the effective $\alpha$ is about 0.1. Since the interaction between the planetary envelope and the disk modifies the rotation rate of the planet, we maintain the forcing of the spin rate of the inner planetary envelope (i.e., $r < 0.96$) during this stage to keep it at the desired value. We find that for the high spin rate model, we have to add angular momentum more aggressively (by a factor of 25 stronger) to maintain steady  planetary envelope rotation. We simulate this second stage for another 800 orbits defined at $r=1$. The final circumplanetary system density profile is shown in the right panel of Figure~\ref{fig:density}. 

\section{Results} \label{sec:results}
We simulate and analyze the circumplanetary systems at different spin rates of the planetary envelope. In \S\ref{subsec:flux}, we calculate the time-averaged radial mass and angular momentum fluxes for different model and demonstrate the existence of a critical spin rate above which mass and angular momentum accretion are prohibited. In \S\ref{subsec:stream}, we present the hydrodynamic flow patterns of the systems. In \S\ref{subsec:tdep}, we discuss the time-dependent accretion. In \S\ref{subsec:high_cs}, we present results for a higher sound speed and demonstrate the dependence of the critical spin rate on this parameter.

\subsection{Mass and angular momentum flux}
\label{subsec:flux}

\begin{figure*}
    \centering
    \hspace*{-0.5cm}
    \includegraphics{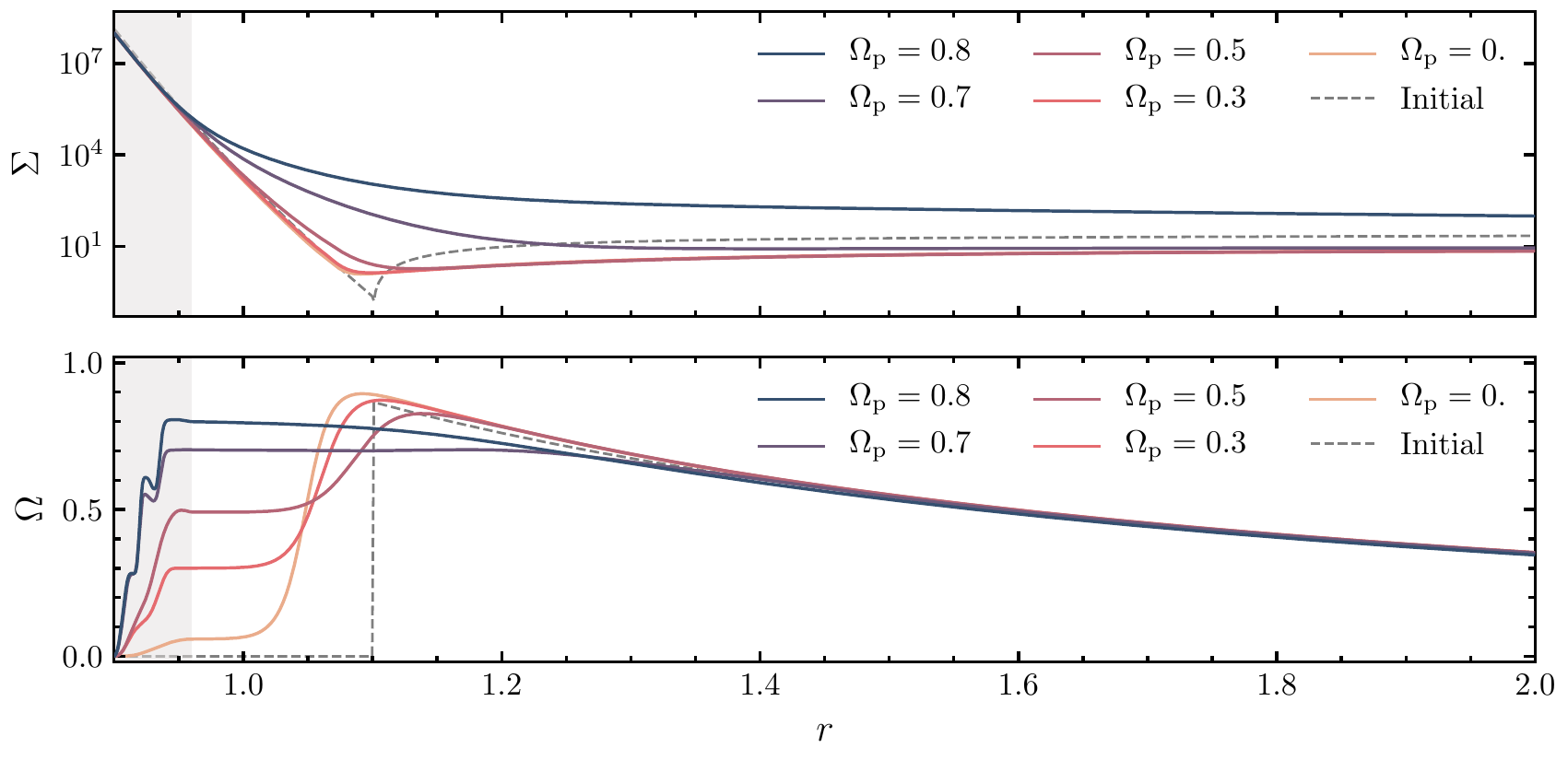}
    \caption{Density profiles integrated over the $\theta$-direction (upper panel) and mid-plane angular velocity profiles (lower panel) at different planetary spin rates,  $\Omega_{\mathrm{p}} \in [0., 0.3, 0.5, 0.7, 0.8]$ at $r=0.96$, for the $c_s=0.1$ setup. The snapshots are taken from the end of the Stage 2 simulations. The grey filled region, $r \in [0.9, 0.96]$, is the nonphysical region in which we force the planetary envelope to keep it rotating at the desired value. The grey dashed lines show the initial conditions, which are the same for all simulations. The density and angular velocity profiles are modified in the steady state. For the low-spin cases ($\Omega_{\mathrm{p}}=0$, 0.3, or 0.5), the outer planetary envelope has a flat angular velocity until it reaches the boundary layer where $\dv*{\Omega}{r} > 0$. For the high-spin cases ($\Omega_{\mathrm{p}}=0.7$ or 0.8), the boundary layer no longer exists. We instead see a flat angular velocity profile for $\Omega_{\mathrm{p}}=0.7$, and a monotonically decreasing profile for $\Omega_{\mathrm{p}}=0.8$. The angular velocity of the outer disk region  follows the Keplerian velocity in all cases.}
    \label{fig:profile}
    
    \bigskip
    
    \hspace*{-0.5cm}
    \includegraphics{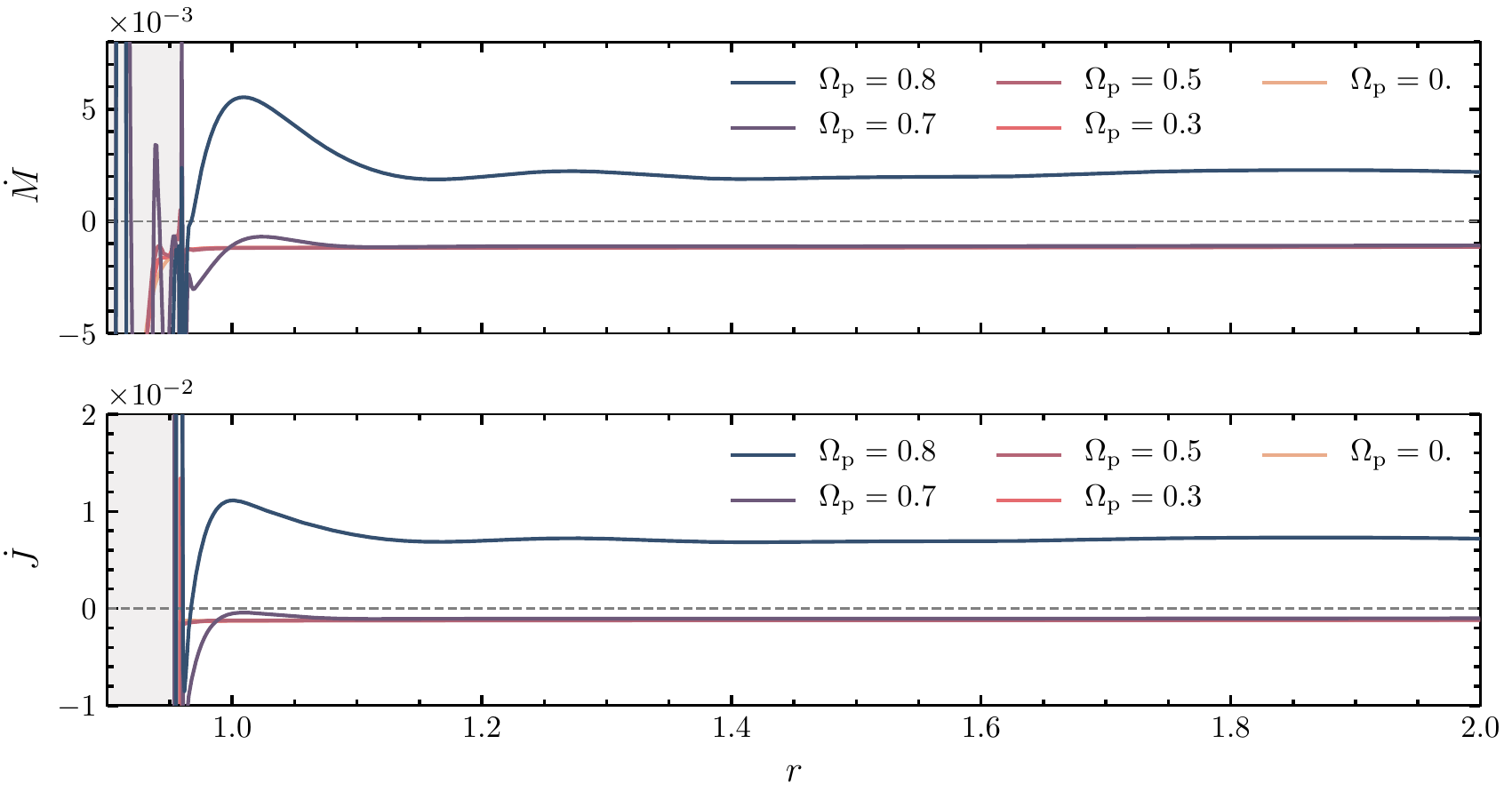}
    \caption{Time-averaged radial mass fluxes (upper panel) and radial angular momentum fluxes (lower panel), averaged over 100 orbits at different planetary spin rates,  $\Omega_{\mathrm{p}} \in [0., 0.3, 0.5, 0.7, 0.8]$ at $r=0.96$, for the $c_s=0.1$ setup. 
    The grey filled region, $r \in [0.9, 0.96]$, is the nonphysical region in which we force the planetary envelope to keep it rotating at the desired value.
    As described in Figure~\ref{fig:profile}, we identify boundary layers (i.e., the region where $\dv*{\Omega}{r} > 0$) at $\Omega_{\mathrm{p}}=0$, 0.3 or 0.5, a flat angular velocity profile at $\Omega_{\mathrm{p}}=0.7$, and a monotonically decreasing angular velocity profile at $\Omega_{\mathrm{p}}=0.8$. Mass and angular momentum are accreted onto the planetary envelope for $\Omega_{\mathrm{p}} \in [0, 0.3, 0.5, 0.7]$, but decreted from the planetary envelope for $\Omega_{\mathrm{p}} = 0.8$. The critical spin rates for effective accretion is between $\Omega_{\mathrm{p}}=0.7$--0.8.}
    \label{fig:flux}
\end{figure*}

We study the properties of the circumplanetary systems at five different spin rates of the central planet ($\Omega_{\mathrm{p}} \in [0., 0.3, 0.5, 0.7, 0.8]$ at $r=0.96$). In Figure~\ref{fig:profile}, we present the density profiles integrated over the $\theta$-direction (upper panel), and the mid-plane angular velocity profiles (lower panel), at the end of the Stage 2 simulations when the inner disk has reached a steady state. The initial conditions of the surface density and angular velocity are shown in grey dashed lines, and are the same for all simulations. 
As shown in Figure~\ref{fig:profile}, the density profiles adjust from the initial setup to balance the planet and disk rotation. The structural adjustment of the outer planetary envelope is most prominent for the high spin-rate cases, $\Omega_{\mathrm{p}} = 0.7$ or 0.8. The planetary rotation in these cases is strong enough to provide significant support against gravity, and the pressure and thus density gradients become correspondingly shallower.
For the low-spin cases ($\Omega_{\mathrm{p}}=0$, 0.3 or 0.5), the final angular velocity profiles are altered by at most a modest amount. 
In the outer disk region, the gas rotates at nearly Keplerian velocity until it approaches the planetary envelope. By continuity, we expect a turning point where $\dv*{\Omega}{r} = 0$ as the near-Keplerian disk matches on to the more slowly rotating planetary envelope, and this turning point is usually defined as the outer radial range of the boundary layer. Throughout the boundary layer, the gas gradually brakes until it matches the rotation rate of the planetary envelope, which is usually defined as the inner radial range of the boundary layer. The gas is then accreted on to the planet. We note that at $\Omega_{\mathrm{p}}=0$, 0.3 or 0.5, the disk rotates at slightly super-Keplerian velocity from $r \sim 1.1$ to 1.4 because that region has a positive density gradient. For the high-spin cases ($\Omega_{\mathrm{p}}=0.7$ or 0.8), the simulated angular velocity profiles are very different from the classical picture. It is no longer obvious how to define the boundary layer. In the $\Omega_{\mathrm{p}}=0.7$ case, we find a flat angular velocity profile and in the $\Omega_{\mathrm{p}}=0.8$ case, we find a monotonically decreasing profile. 

To understand how the change of the angular velocity profiles affects the accretion, we calculate the time-averaged radial mass and angular momentum flux. The radial mass flux $\dot{M}$ can be expressed as
\begin{equation}
    \dot{M} = \iint \rho v_r r^2 \sin{\theta} d\phi d\theta,
\end{equation}
where $\rho$ and $v_r$ are the density and radial velocity, respectively.
In our 2-dimensional, axisymmetric setup, we compute the radial mass flux at cell $(r_i, \theta_j)$ as,
\begin{equation}
    \dot{M}(r_i, \theta_j) = 2\pi \rho v_r(r_i, \theta_j) r_i^2(\cos{\theta_{j-1/2}-\cos{\theta_{j+1/2}}}).
\end{equation}
A negative $\dot{M}$ indicates gas accretion onto the planet from the disk, whereas a positive $\dot{M}$ indicates gas \emph{decretion} from the planet to the disk.
To calculate the radial angular momentum transport, we write down the angular momentum equation \citep[Eq.~4 in][]{kley93},
\begin{equation} \label{eqn:mom}
    \frac{\partial (\rho r v_\phi \sin{\theta})}{\partial t} + \div(\rho r v_\phi \sin{\theta} \vb*{v} - r\sin{\theta}\vb*{t_\phi}) = 0,
\end{equation}
where $\vb*{t_\phi}$ is the $\phi$-component of the viscous stress tensor, written as $\vb*{t_\phi} = (t_{\phi r},t_{\phi \theta},t_{\phi \phi})$. Specifically, the tensor component in the $r$-direction is $t_{\phi r} = \rho \nu r\sin{\theta} ({\partial \Omega}/{\partial r})$, where $\Omega$ is the angular velocity. We expect zero net angular momentum flux in the $\theta$-direction because our simulation grid extends from $\theta=0$ to $\theta=\pi$. Plugging in the tensor component $t_{\phi r}$, we may rewrite Equation (\ref{eqn:mom}) as,
\begin{equation} \label{eqn:mom_rewrite}
    \frac{\partial (\rho r v_\phi \sin{\theta})}{\partial t} + \frac{1}{r^2}\frac{\partial}{\partial r}\Big(\rho r^3 v_r v_\phi \sin{\theta} - \rho \nu r^4 \sin^2{\theta} \frac{\partial \Omega}{\partial r}\Big) = 0,
\end{equation}
where the first term on the left hand side is the rate of angular momentum change per unit volume, the first term in the parentheses describes the radial angular momentum change due to the advection, and the second term in the parentheses describes the viscous torque. In a steady state, the parenthesized terms should be a constant as a function of the radial distance. To calculate the radial angular momentum flux $\dot{J}$, we integrate the parenthesized terms by $\int d\Omega$ where $d\Omega = \sin{\theta}d\phi d\theta$,
\begin{multline}
    \dot{J} = \iint \rho r^3 v_r v_\phi \sin^2{\theta} d\phi d\theta + \iint -\rho \nu r^4 \sin^3{\theta} \frac{\partial \Omega}{\partial r} d\phi d\theta.
\end{multline}
To apply the equation to our simulation grid, we approximate $\dot{J}$ at cell $(r_i, \theta_j)$ as,
\begin{align}
    \dot{J}(r_i, \theta_j) &= 2\pi \rho r_i^3  v_r v_\phi \big[\frac{1}{2}(\theta_{j+1/2}-\theta_{j-1/2}) \nonumber\\
    &\qquad \qquad \qquad \qquad -\frac{1}{4}(\sin{2\theta_{j+1/2}}-\sin{2\theta_{j-1/2})}\big] \nonumber\\
    &\quad -2\pi \rho \nu r_i^4 \frac{\Omega_{i+1,j}-\Omega_{i,j}}{r_{i+1}-r_i}
    \big[(\cos{\theta_{j-1/2}-\cos{\theta_{j+1/2}}}) \nonumber\\
    &\qquad \qquad \qquad \qquad -\frac{1}{3}(\cos^3{\theta_{j-1/2}-\cos^3{\theta_{j+1/2}}})\big]
\end{align}
where $\rho$, $v_r$, and $v_\phi$ are all functions of $(r, \theta)$.

In Figure~\ref{fig:flux}, we present the radial mass flux (upper panel) and radial angular momentum flux (lower panel) averaged over 100~orbits at $r = 1$ at the end of Stage 2 simulations. The inner circumplanetary systems have reached a steady state, illustrated by the  constant fluxes beyond $r = 1.1$. The flat profiles extend up to $r \sim 2.5$. For $\Omega_{\mathrm{p}} = 0$, 0.3, 0.5 or even 0.7, the mass and angular momentum fluxes are negative and the central planet is accreting mass and angular momentum to grow and spin up through its disk. We find the fluxes have similar values for these spin rates, implying that there is no strong dependence on planetary spin rates as long as accretion occurs. The system at these spin rates is controlled by the {\em supply} of mass and angular momentum from the circumplanetary disk, and the precise nature of the boundary condition the planet presents at the inner edge of the boundary layer is unimportant. 
At $\Omega_{\mathrm{p}} = 0.8$, conversely, both the mass and the angular momentum flux become positive. At this spin rate the planet is \emph{decreting} mass and losing angular momentum,  leading to long-term mass loss and spin down. 
From these results, we infer that for this sound speed there exists a critical spin rate in the region between $\Omega_{\mathrm{p}} = 0.7$ and $\Omega_{\mathrm{p}} = 0.8$. Beyond the critical spin rate, the planet would no longer grow and spin up through its disk, but will instead lose mass and spin down. Comparing with the angular velocity profiles in Figure~\ref{fig:profile}, the critical spin rate corresponds to where the flat angular velocity profile switches to a monotonically decreasing angular velocity profile.  

\subsection{Gas flow streamlines}
\label{subsec:stream}

\begin{figure*}
    \centering
    \includegraphics[width=7in]{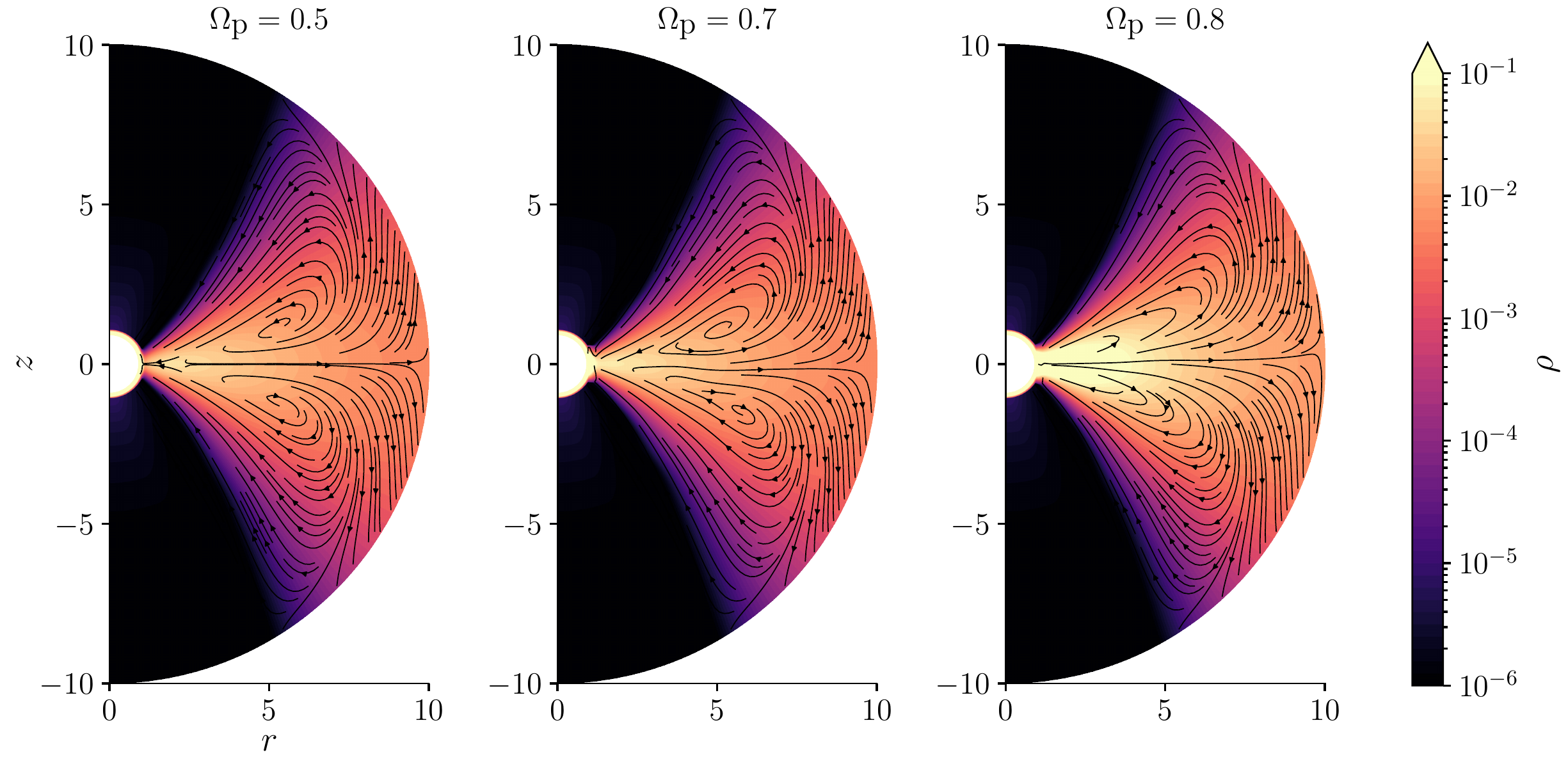}
    \caption{Density profile over-plotted with time-averaged gas flow patterns at representative spin rates for the $c_s=0.1$ setup. The flow patterns are averaged over 100 orbits at $r=1$. Meridional circulation patterns---characterized by a radial outflow at and near the midplane with a radial inflow at high latitudes---are found at all spin rates. At $\Omega_{\mathrm{p}} = 0.5$ and 0.7, we find radial inflow at the mid-plane in the inner disk region where $r<2$. At $\Omega_{\mathrm{p}} = 0.8$, we find radial outflow at the mid-plane at all radii.}
    \label{fig:streamline}

    \bigskip
    
    \hspace*{-1cm}
    \includegraphics[scale=0.9]{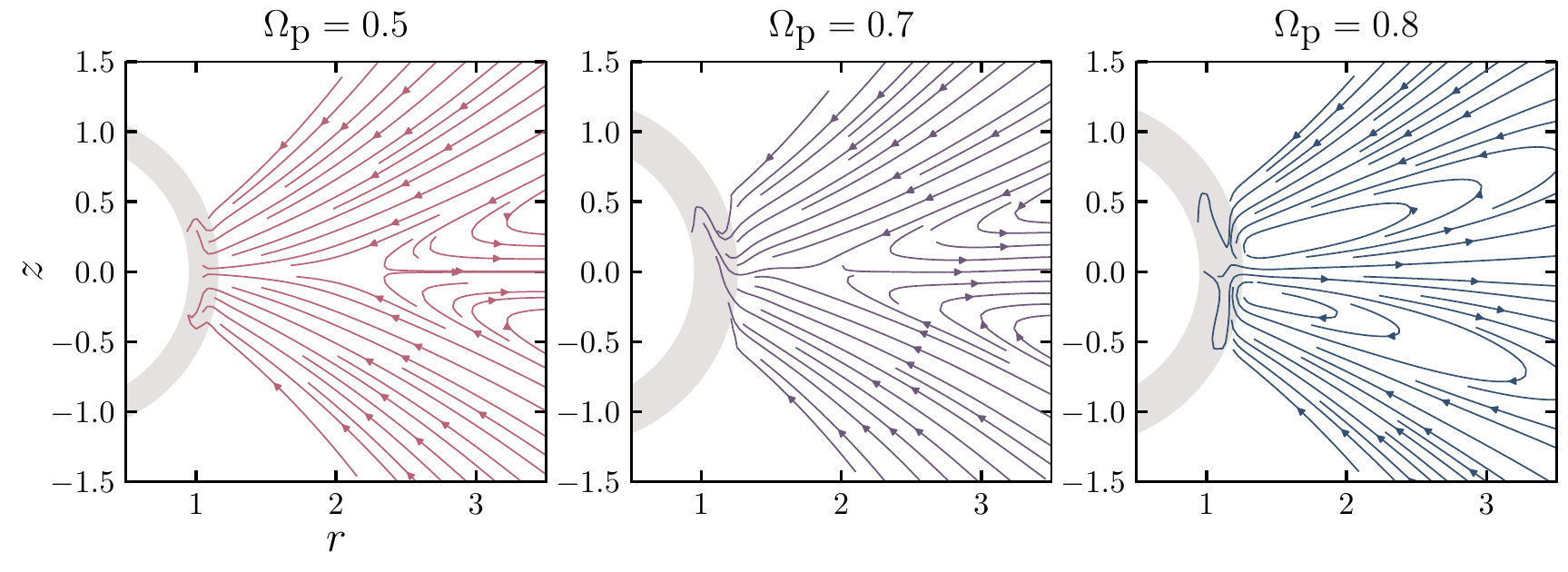}
    \caption{Time-averaged gas flow patterns near the boundary layer at representative spin rates for the $c_s=0.1$ setup. The grey shaded region represents the planetary envelope and boundary layer. For high spin cases with no defined boundary layer, we set $r=1.2$ as the outer boundary. At $\Omega_{\mathrm{p}} = 0.5$, gas from all disk latitudes flows inwards and accretes onto the planetary envelope to high latitude. The radial inflow at the mid-plane changes direction near $r=2$ to radial outflow. At $\Omega_{\mathrm{p}} = 0.7$, the disk flow pattern is similar to the $\Omega_{\mathrm{p}} = 0.5$ case, but gas can only accrete onto part of the planetary envelope due to the fast rotation of the planet. Finally, at $\Omega_{\mathrm{p}} = 0.8$, gas flows outwards from the planetary envelope to the disk.}
    \label{fig:streamline_zoomin}
\end{figure*}

\begin{figure}
    \centering
    \hspace*{-1.5cm}
    \includegraphics[scale=1.1]{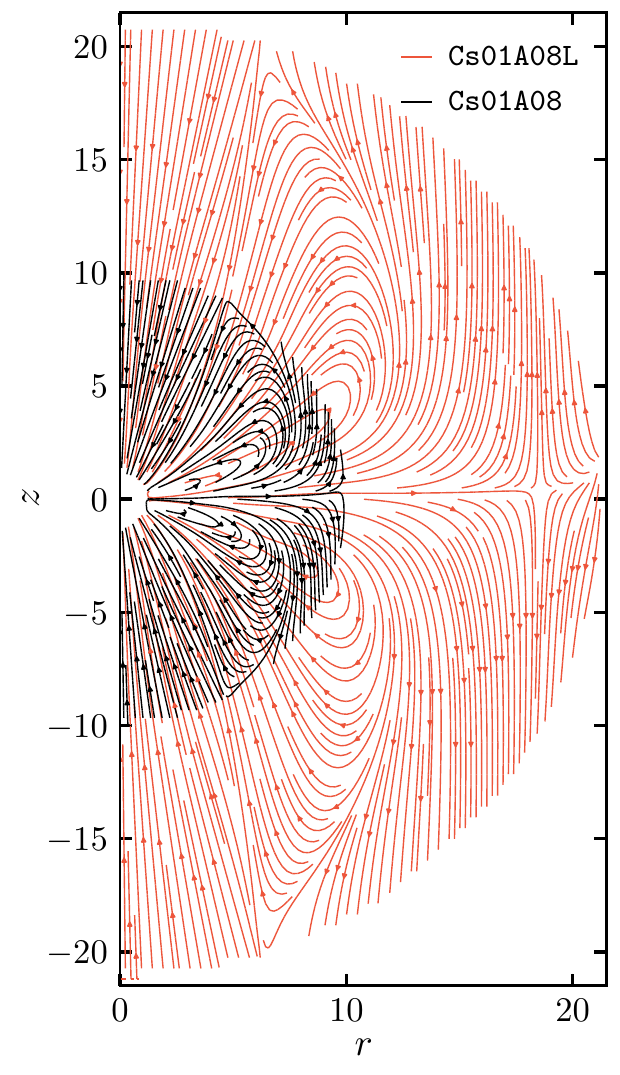}
    \caption{Streamline plot of the large disk simulation ($\mathtt{Cs01A08L}$, in orange) over-plotted with the regular disk simulation  ($\mathtt{Cs01A08}$, in black). The large disk simulation reproduces the meridional circulation flow at a similar radial distance as the one from the regular disk simulation, indicating that the meridional circulation patterns found in the regular disk simulation are not introduced by the outer boundary condition in the $r$-direction.}
    \label{fig:large_disk}
\end{figure}

An advantage of our 2-dimensional simulations is that we are able to capture gas flow patterns at different disk latitudes. In Figure~\ref{fig:streamline}, we show the {\em time-averaged} gas streamlines at representative spin rates. The three spin rates we choose represent an accreting planet with a boundary layer (the $\Omega_{\mathrm{p}} = 0.5$ case), an accreting planet with a flat angular velocity profile (the $\Omega_{\mathrm{p}} = 0.7$ case), and a \emph{decreting} planet with a monotonically decreasing angular velocity profile (the $\Omega_{\mathrm{p}} = 0.8$ case). The flow patterns are similar for all the low-spin cases, so we only present the $\Omega_{\mathrm{p}} = 0.5$ case for illustration. At all spin rates, we observe a meridional circulation pattern which can be described as a radial outflow at and near the midplane with a radial inflow at high latitudes. Such a pattern was predicted for an isothermal, viscous accretion disk in previous work \citep[e.g.,][]{urpin84, philippov17}. Although the gas is transported outwards at the midplane, the boundary layer and the inner disk can still have a net mass inflow if more mass can be transported inwards from high latitudes. At $\Omega_{\mathrm{p}} = 0.5$ and 0.7, we find a radial inflow at the midplane in the inner disk region extending from $r=1$ to $r=2$; whereas at $\Omega_{\mathrm{p}} = 0.8$, we find a radial outflow at the midplane even for the inner disk region. The radial flow direction we observed is consistent with the prediction of \cite{philippov17}. For the $\Omega_{\mathrm{p}} = 0.7$ case, the circulation patterns are slightly asymmetric about the mid-plane. Since the outer disk has not yet reached to a steady state during our simulation timescale, the asymmetry pattern is not too surprising.

Our high resolution simulations allow us to resolve flow patterns near the boundary layer. In Figure~\ref{fig:streamline_zoomin}, we present the detailed flow patterns in the outer planetary envelope and the boundary layer. At $\Omega_{\mathrm{p}} = 0.5$, gas from all disk latitudes flows inwards and accretes onto the planetary envelope first at the equator and then to high latitude. The midplane radial velocity $\abs{v_r}$ increases sharply to a maximum of $\sim$ 40\% of the sound speed as the gas approaches to the boundary layer region and then decreases sharply in the boundary layer to accrete onto the planetary surface. The radial inflow at the midplane changes its direction near $r=2$ to radial outflow. At $\Omega_{\mathrm{p}} = 0.7$, the disk flow pattern is similar to the $\Omega_{\mathrm{p}} = 0.5$ case, but gas can only accrete onto part of the planetary envelope due to the fast rotation of the planetary envelope. The maximum $\abs{v_r}$ is $\sim$ 6\% of the sound speed, which is about an order of magnitude smaller than the $\Omega_{\mathrm{p}} = 0.5$ case. Lastly, at $\Omega_{\mathrm{p}} = 0.8$, gas can no longer accrete onto the planetary envelope. Most of the gas flows outwards from the planetary envelope to the disk causing \emph{decretion} and spin-down of the planetary envelope.

The meridional circulation is a large-scale flow structure which extends throughout the simulated circumplanetary disk. To confirm that the circulation pattern in the inner disk, near the boundary layer, is not an artificial feature introduced by the outer boundary conditions, we ran an additional simulation of the $\Omega_{\mathrm{p}} = 0.8$ case with an extended radial range out to $r=20$. We keep the spatial resolution and other parameters the same. As shown in Figure~\ref{fig:large_disk}, the meridional circulation pattern is again observed in the large disk simulation ($\mathtt{Cs01A08L}$) and is located at a similar location and latitude to the regular disk simulation ($\mathtt{Cs01A08}$). This indicates that the observed meridional flow features in the inner disk are not introduced by the boundary condition.

\subsection{Time-dependent variations}
\label{subsec:tdep}

\begin{figure*}
    \centering
    \hspace*{-0.5cm}
    \includegraphics[scale=1]{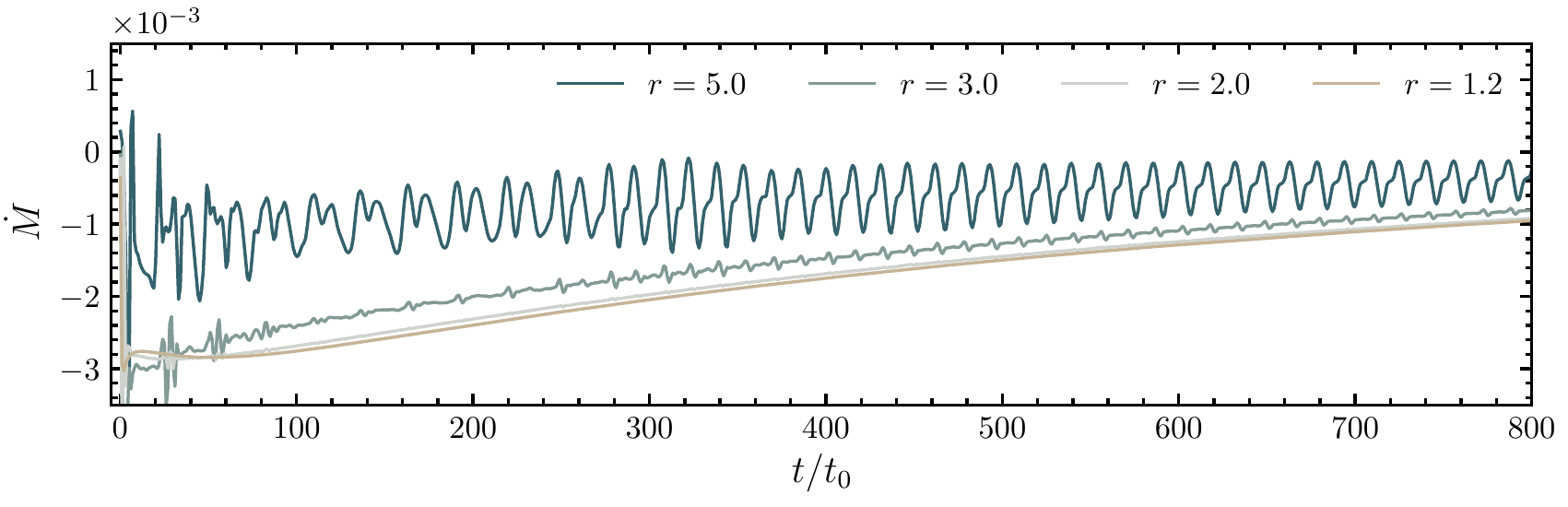}
    \caption{Time-dependent radial mass fluxes at different radial distances during the Stage 2 simulation for the $c_s = 0.1$, $\Omega_{\mathrm{p}}=0.3$ case. $\dot{M}$ is calculated at intervals of $t_0$, defined as one orbit at $r=1$. Periodic variations in the mass flux are observed at various radial distances. The small amplitude variations may be caused by the variability of the meridional flow. The inner disk region (i.e., $r < 2.5$) has reached a steady state after $\sim 600$ orbits and has a constant $\dot{M}$ over this radial range (e.g., the upper panel of Figure~\ref{fig:flux} and the overlapping of the orange and yellow curves in this figure). The outer disk has not yet reached a steady state during the simulation timescale and its instantaneous and time-averaged $\dot{M}$ deviate from the inner disk value.}
    \label{fig:mdot_tdep}
\end{figure*}

We presented the time-averaged radial mass and angular momentum fluxes, and the averaged gas flow streamlines, in previous subsections. These properties also display time-dependent features introduced by the variability of the meridional flows. To demonstrate the variability, we plot the radial mass fluxes as a function of time at different radial distances in Figure~\ref{fig:mdot_tdep}. The data is taken from Stage 2 of our simulation (i.e., after adding in the kinematic viscosity) for the $c_s = 0.1$, $\Omega_{\mathrm{p}}=0.3$ case. We present the $\Omega_{\mathrm{p}}=0.3$ case because the variation of the meridional circulation pattern is most obvious for the $\Omega_{\mathrm{p}}=0$ and 0.3 cases. The radial mass flux $\dot{M}$ is calculated every $t_0$ which is defined as one orbit at $r = 1$. As shown in Figure~\ref{fig:mdot_tdep}, the radial mass flux varies periodically, both in the inner disk (e.g. at $r = 3$) and the outer disk (e.g. at $r \ge 5$). The amplitude of the radial flux variation is greater in the outer disk, indicating a stronger meridional flow variation.  Figure~\ref{fig:mdot_tdep} also illustrates the response of the circumplanetary system to the addition of disk viscosity. The inner disk region ($r < 2.5$) has reached a steady state after $\sim$ 600 orbits and has a constant $\dot{M}$ over that radial range (see, e.g., the upper panel of Figure~\ref{fig:flux} and the overlapping of the orange and yellow curves). The outer disk has not yet reached a steady state during our simulation timescale (800 orbits at $r = 1$) and its instantaneous and time-averaged $\dot{M}$ deviate from the inner disk value.

\begin{figure*}
    \centering
    \hspace*{0cm}
    \includegraphics[scale=0.8]{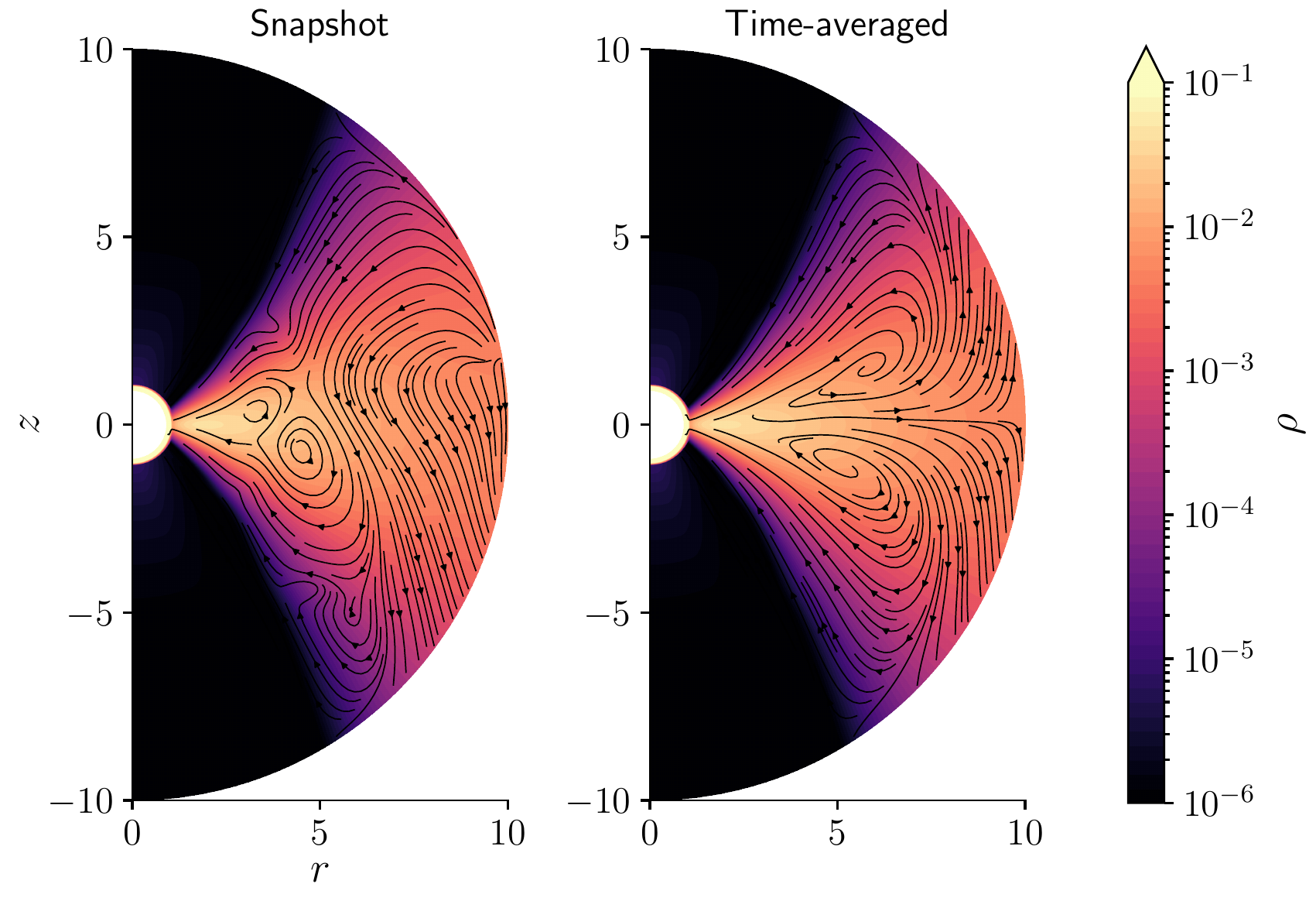}
    \caption{An instantaneous (left panel) versus the time-averaged (right panel) gas flow streamlines for the $c_s = 0.1$, $\Omega_{\mathrm{p}}=0.3$ case. The left panel presents a snapshot of the gas flow pattern which has several regional and small scale meridional circulations that are not symmetric around the mid-plane. The right panel presents the flow pattern averaged over 100 orbits defined at $r=1$. The local turbulent flow patterns observed in the left panel are averaged out. The meridional flow is unstable and shows variation at short timescales.}
    \label{fig:streamline_tdep}
\end{figure*}

The variation of the meridional flows can also be visualized via streamline plots. In Figure~\ref{fig:streamline_tdep}, we compare a snapshot of the gas flow streamlines (the left panel) to the time-averaged gas flow streamlines (the right panel), again for the $c_s = 0.1$, $\Omega_{\mathrm{p}}=0.3$ case. The time-averaged flow pattern is similar to the one of the $c_s = 0.1$, $\Omega_{\mathrm{p}}=0.5$ case (shown in the left panel of Figure~\ref{fig:streamline}). As shown in the left panel of Figure~\ref{fig:streamline_tdep}, several regional and small scale meridional circulation patterns are found. These circulation patterns are unstable and appear at different locations at different times. When we integrate the system for a long timescale, the turbulent features average out and we identify the steady streamline pattern shown in the right panel of Figure~\ref{fig:streamline_tdep}.
The meridional flow induces vertical velocities that are 1--2 \% of the sound speed. The time-dependent nature of these flows, and the fact that there is a non-zero instantaneous velocity at $z=0$, means that they would oppose dust settling if present in real circumplanetary disks. We consider solid particles with material density $\rho_{\rm m}$ and radius $s$, interacting aerodynamically with gas of density $\rho$ and thermal speed $v_{\rm th}$. The settling velocity in the Epstein drag regime (appropriate to most circumplanetary disk conditions) is \citep{armitage10},
\begin{equation}
    v_{\rm settle} = \frac{\rho_{\rm m}}{\rho} \frac{s}{v_{\rm th}} \Omega^2 z.
\end{equation}
The condition for time-dependent meridional flows with characteristic velocity $\epsilon c_s$ to fully mix particles can be estimated by requiring that,
\begin{equation}
    v_{\rm settle} < \epsilon c_s,
\end{equation}
at $z=h$. Dropping numerical factors that are of the order of unity, this condition can be written as,
\begin{equation}
    \frac{\rho_{\rm m} s}{\Sigma} \lesssim \epsilon,
\end{equation}
where $\Sigma$ is the circumplanetary disk gas surface density. For mm-sized icy particles ($\rho_{\rm m} \approx 1 \ {\rm g \ cm^{-3}}$), and $\epsilon \approx 0.01$, we then estimate that meridional flows would be strong enough to oppose settling in disks with $\Sigma \lesssim 10 \ {\rm g \ cm^{-2}}$. This is a relatively relaxed condition. It implies that settling would not occur for most observationally interesting particle sizes ($s \sim {\rm mm}$, or smaller) provided that the circumplanetary disk is at least moderately optically thick (assuming standard opacities and gas-to-dust ratios).

\subsection{The dependence on isothermal sound speed}
\label{subsec:high_cs}
\begin{figure*}
    \centering
    \hspace*{-0.5cm}
    \includegraphics{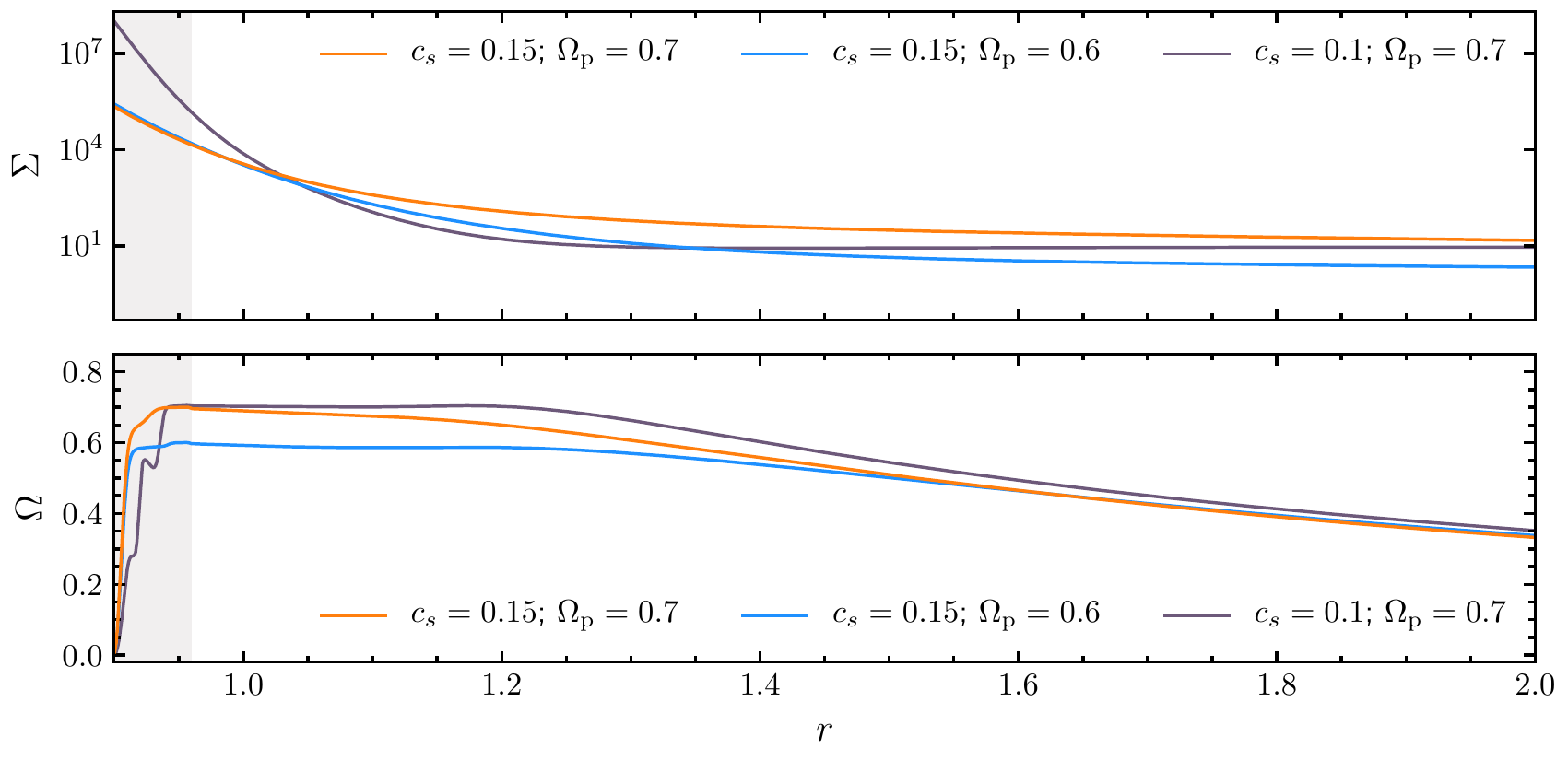}
    \caption{Density profiles integrated over the $\theta$-direction (upper panel) and mid-plane angular velocity profiles (lower panel) at different planetary spin rates,  $\Omega_{\mathrm{p}} \in [0.6, 0.7]$ at $r=0.96$, for the $c_s=0.15$ setup. The snapshots are taken from the end of the Stage 2 simulations. The grey filled region, $r \in [0.9, 0.96]$, is the nonphysical region in which we force the planetary envelope to keep it rotating at the desired value. The boundary layer does not exist for either spin rate. We observe a flat angular velocity profile for $\Omega_{\mathrm{p}}=0.6$ and a monotonically decreasing profile for $\Omega_{\mathrm{p}}=0.7$. For comparison, we plot the $\Omega_{\mathrm{p}}=0.7$ curve in Figure~\ref{fig:profile} for the $c_s=0.1$ setup, where a flat angular velocity profile was found.}
    \label{fig:profile_cs}

    \bigskip
    
    \hspace*{-0.5cm}
    \includegraphics{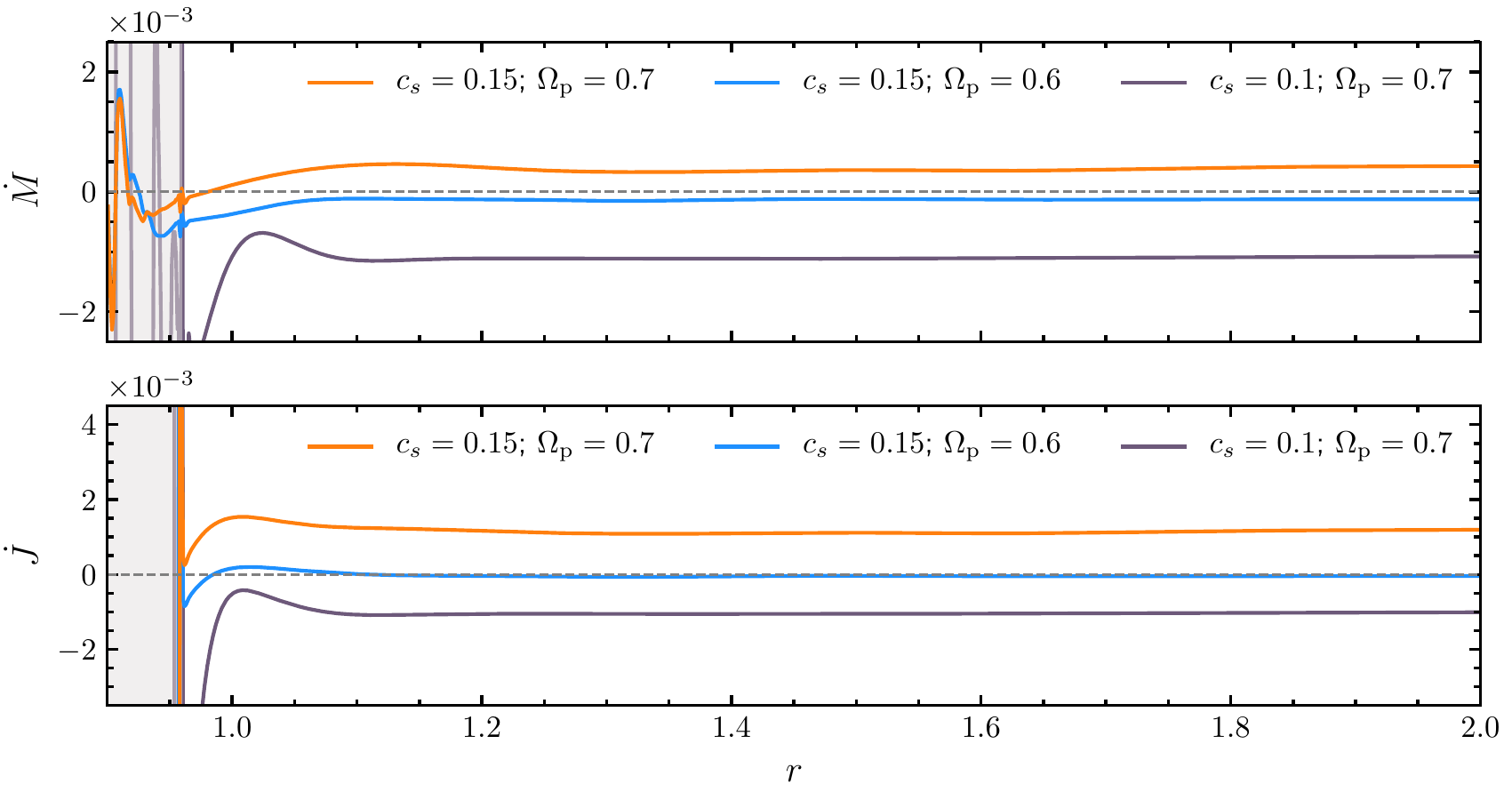}
    \caption{Time-averaged radial mass fluxes (upper panel) and radial angular momentum fluxes (lower panel) taken from the end of Stage 2 simulations averaged over 100 orbits at different planetary spin rates, $\Omega_{\mathrm{p}} \in [0.6, 0.7]$ at $r=0.96$, for the $c_s=0.15$ setup. The grey filled region, $r \in [0.9, 0.96]$, is the nonphysical region in which we force the planetary envelope to keep it rotating at the desired value. Mass and angular momentum are accreted onto the planetary envelope at $\Omega_{\mathrm{p}} = 0.6$, but decreted from the planetary envelope at $\Omega_{\mathrm{p}} = 0.7$. This indicates a critical spin rate between $\Omega_{\mathrm{p}}=0.6-0.7$ for an isothermal sound speed of 0.15. For comparison, we plot the $\Omega_{\mathrm{p}} = 0.7$ curve in Figure~\ref{fig:flux} for the $c_s=0.1$ setup, which has a critical spin rate between 0.7--0.8.}
    \label{fig:flux_cs}
\end{figure*}

We now examine the dependence of the critical spin rate on the isothermal sound speed. In previous simulations, we adopted an isothermal sound speed of 0.1 and found that the critical spin rate was between 0.7--0.8 of the breakup angular velocity. Here, we set an isothermal sound speed of 0.15 and evaluate the resulting critical spin rate. Given the experience from previous simulations, we run simulations at two spin rates, $\Omega_{\mathrm{p}}=0.6$ and $\Omega_{\mathrm{p}}=0.7$. In Figure~\ref{fig:profile_cs}, we show their density and angular velocity profiles. In Figure~\ref{fig:flux_cs}, we show their radial mass and angular momentum fluxes. At $\Omega_{\mathrm{p}}=0.6$, the system has a flat angular velocity profile and both the radial mass flux and the angular momentum fluxes are negative, whereas at $\Omega_{\mathrm{p}}=0.7$, the system has a monotonically decreasing angular velocity profile and both fluxes are positive. Our simulations thus indicate a critical spin rate between 0.6--0.7 of the planet's breakup angular velocity for $c_s = 0.15$. 

We find a dependence of critical spin rate on the isothermal sound speed of the circumplanetary system. The critical spin rate decreases as the isothermal sound speed increases. Since the width of boundary layer roughly scales as $c_s^2$ \citep{lynden-bell74, pringle77}, the higher sound speed leads to a wider boundary layer. The critical spin rate corresponds to the condition where the boundary layer no longer exists, i.e., an infinitely wide boundary layer. The critical condition is easier to achieve at a higher sound speed, and therefore we find a lower critical spin rate for a higher isothermal sound speed system.

\section{Discussion and Conclusion} \label{sec:conclusion}

In this work we have made an estimate of the terminal rotation rate of unmagnetized planets, which accrete gas from their circumplanetary disks via an equatorial boundary layer. As prior work on boundary layers \citep{popham91} makes clear, the limiting spin in this mode of accretion---although rapid---falls short of breakup. Rather, an equilibrium state can be reached in which the accreted specific angular momentum equals that of the central object, and spin-up ceases. In our model system, where the planet, boundary layer and disk are taken to be a viscous isothermal fluid, with axisymmetry assumed, we find that the equilibrium spin rate is between 70\% and 80\% of breakup if the disk aspect ratio near the planet is $h/r=0.1$ (Figure~\ref{fig:profile} and \ref{fig:flux}). The equilibrium spin drops to between 60\% and 70\% for a thicker disk, with $h/r=0.15$ (Figure~\ref{fig:profile_cs} and \ref{fig:flux_cs}; see Table~\ref{tbl:sims} for a summary). The boundary layer and circumplanetary disk are found to be variable, on time scales of the order of ten inner orbital periods, due to the development of meridional flows within the circumplanetary disk \citep[e.g., Figure~\ref{fig:streamline}, \ref{fig:streamline_zoomin}, and \ref{fig:mdot_tdep};][]{urpin84,kley92,fromang11,philippov17}. The variability arises because of time-dependence introduced by the meridional flow itself, which over time loses reflection symmetry across the disk equator (Figure~\ref{fig:streamline_tdep}). \citet{popham91}, in the context of compact object accretion, demonstrated that boundary layer accretion leads to sub-break up central object spin. We confirm their key finding, but note that our results differ from theirs at a qualitative level. Specifically, whereas \citet{popham91} obtained solutions in which the central object accreted mass while spinning down, our spin-down solutions are accompanied by {\em decretion}. This difference is likely to be a consequence of contrasting formulations of the problem. \citet{popham91} solved directly for the steady-state solution of one-dimensional, vertically averaged, disk equations. This approach leads to a boundary value problem, in which the specific angular momentum of the flow $\dot{j}$ is recovered for {\em specified} $\dot{M} > 0$ as an eigenvalue. By dropping the vertical averaging and moving to axisymmetry, we instead solve for the quasi-steady state of an initial value problem, applying boundary conditions deep in the planetary interior, and far out in the disk, that are intended to be minimally coercive to the boundary layer properties. With this approach, we find that the transition between spin-up and spin-down occurs at the same (or very similar) spin rate as that between accretion and decretion. Our results agree with those of \citet{hertfelder17}, who also found decretion at high spin rates in radiation hydrodynamics simulations of compact object boundary layers.

\begin{deluxetable*}{lccccccrr}
\tablecaption{Simulation runs and a summary of circumplanetary system properties \label{tbl:sims}}
\tablewidth{6pt}
\tablehead{
\colhead{Run} & \colhead{$c_s$} & \colhead{$\Omega_{\mathrm{p}}$} & \colhead{$r$ range, $N_r$} & \colhead{$\theta$ range, $N_\theta$} & \colhead{BL Slope} & \colhead{$\rho (r=1.2)$} & \colhead{$\dot{M} (r=1.2)$} & \colhead{$\dot{J} (r=1.2)$}}
\startdata
$\mathtt{Cs01A0}$ & 0.1 & 0.0 & [0.9, 10], 2048 & [0, $\pi$], 2048 & $+$ & $3.20 \times 10^{0}$ & $-1.17 \times 10^{-3}$ & $-1.22 \times 10^{-3}$ \\
$\mathtt{Cs01A03}$ & 0.1 & 0.3 & [0.9, 10], 2048 & [0, $\pi$], 2048 & $+$ & $3.13 \times 10^{0}$ & $-1.17 \times 10^{-3}$ & $-1.22 \times 10^{-3}$ \\
$\mathtt{Cs01A05}$ & 0.1 & 0.5 & [0.9, 10], 2048 & [0, $\pi$], 2048 & $+$ & $3.06 \times 10^{0}$ & $-1.18 \times 10^{-3}$ & $-1.23 \times 10^{-3}$ \\
$\mathtt{Cs01A07}$ & 0.1 & 0.7 & [0.9, 10], 2048 & [0, $\pi$], 2048 & 0 & $1.74 \times 10^{1}$ & $-1.11 \times 10^{-3}$ & $-1.05 \times 10^{-3}$ \\
$\mathtt{Cs01A08}$ & 0.1 & 0.8 & [0.9, 10], 2048 & [0, $\pi$], 2048 & $-$ & $3.91 \times 10^{2}$ & $1.98 \times 10^{-3}$ & $6.96 \times 10^{-3}$ \\
$\mathtt{Cs01A08L}$ & 0.1 & 0.8 & [0.9, 21.2], 2688 & [0, $\pi$], 2048 & $-$ & $3.37 \times 10^{2}$ & $6.37 \times 10^{-5}$ & $4.26 \times 10^{-3}$ \\
\hline
$\mathtt{Cs015A06}$ & 0.15 & 0.6 & [0.9, 10], 2048 & [0, $\pi$], 2048 & 0 & $3.50 \times 10^{1}$ & $-1.18 \times 10^{-4}$ & $-3.77 \times 10^{-5}$ \\
$\mathtt{Cs015A07}$ & 0.15 & 0.7 & [0.9, 10], 2048 & [0, $\pi$], 2048 & $-$ & $1.20 \times 10^{2}$ & $4.10 \times 10^{-4}$ & $1.16 \times 10^{-3}$ \\
\enddata
\tablecomments{BL, boundary layer; $\dot{M}, \dot{J}$ are in $r$-direction; $\dot{M}<0$ indicates mass accretion onto the planet and $\dot{J}<0$ indicates angular momentum transport onto the planet. See \S\ref{sec:results} for more details.}
\end{deluxetable*}

Our results suggest that, if some subset of young giant planets lack strong dipolar magnetic fields, they would spin up during the accretion phase to between 60\% and 80\% of their break-up speed. Subsequent spin-up, occurring at constant angular momentum as the planets contract, is strongly mass-dependent \citep{fortney11}. Planets with masses between those of Saturn and Jupiter contract substantially, but on a time scale that is shorter at higher masses (accordingly, any early braking process works better for Jupiter than for Saturn). Ice giant contraction is much less significant. At late times, exoplanets that were experienced boundary layer accretion during their growth phase would then be expected to spin with a {\em minimum} of 60\% of their break-up speed. Such planets might be oblate enough to identify from transit data \citep{seager02}, or from future spectroscopic observations. Boundary layer and magnetospheric accretion \citep{batygin18} also differ in their predictions for protoplanetary properties. Magnetospheric accretion results in accretion shocks and (typically) localized hot spots on the planetary surface, leading to photometric modulation on the planetary spin period and strong emission in lines such as H$\alpha$. A boundary layer is not expected to be a strong source of line emission, and will only produce a thermal component that is distinguishable from that of the disk if the planet is slowly spinning. Non-axisymmetric boundary layer instabilities---which are not captured in our simulations---may lead to variability with a time scale comparable to that of the Keplerian orbital period at the surface of the central object \citep{belyaev12}. Variability on these time scales that may be associated to the boundary layer is observed in dwarf nova systems \citep{warner04}. We note that if cirumplanetary disk accretion is strongly episodic, as suggested by \citet{brittain20}, then boundary layer accretion during high accretion rate phases could co-exist with magnetospheric accretion during quiescence.

Future work will need to address two obvious limitations of the present study. First, by adopting an isothermal equation of state, we do not capture the often-substantial release of energy in the boundary layer. The dynamical effects of that energy release---in the form of a thickening and radial broadening of the boundary layer region---can be modeled using viscous radiation hydrodynamics simulations, as was already done for protostellar systems by \citet{kley96} and, more recently, for compact objects by \citet{hertfelder17}. We cannot exclude the possibility that such thermal effects could broaden protoplanetary boundary layers substantially, thereby reducing the predicted terminal spin rate. Second, our assumption of axisymmetry means that we cannot attempt to represent physical mechanisms of angular momentum transport within either the disk or the boundary layer. Three-dimensional simulations \citep[e.g.][]{philippov16,belyaev18} that model the transport processes properly are needed to predict the detailed structure of protoplanetary boundary layers, along with potentially observable properties such as their intrinsic variability.

\acknowledgments
JD gratefully acknowledges support and hospitality from the pre-doctoral program at the Center for Computational Astrophysics, Flatiron Institute. Research at the Flatiron Institute is supported by the Simons Foundation. JD is supported in part by the Shaffer Career Development fund. PJA acknowledges support from NASA TCAN award 80NSSC19K0639.
The Center for Exoplanets and Habitable Worlds is supported by the Pennsylvania State University, the Eberly College of Science, and the Pennsylvania Space Grant Consortium. 

\software{$\mathtt{Athena}$++ \citep{stone20}, $\mathtt{Matplotlib}$ \citep{hunt07, droe16}, $\mathtt{Numpy}$ \citep{vand11, harr20}, $\mathtt{Jupyter}$ \citep{kluy16}}


\bibliographystyle{aasjournal}
\bibliography{planet}

\end{CJK*}
\end{document}